\documentclass[reprint,english,aps,prb,twocolumn,amsmath, amssymb,
showpacs,showkeys]{revtex4} \usepackage[T1]{fontenc}
\usepackage{subfigure} \usepackage[latin1]{inputenc}
\usepackage{graphicx} \usepackage{epstopdf} \usepackage{epsfig}
\usepackage{prettyref} \usepackage{babel} \usepackage{color}
\makeatother

\begin{document}

\title{Spin canted magnetism, decoupling of charge and spin ordering in
NdNiO$_3$}

\author{Devendra Kumar}
\email{deveniit@gmail.com}\altaffiliation{Present Address: UGC-DAE Consortium for Scientific Research, University Campus, Khandwa Road, Indore-452001, India.}\affiliation{Department of Physics, Indian
Institute of Technology Kanpur 208016, India}

\author{K. P. Rajeev} \email{kpraj@iitk.ac.in}\affiliation{Department of
Physics, Indian Institute of Technology Kanpur 208016, India}

\author{J. A. Alonso} \affiliation{Instituto de Ciencia de Materiales de
Madrid, CSIC, Cantoblanco, E-28049 Madrid, Spain}

\author{M. J. Mart\'inez-Lope} \affiliation{Instituto de Ciencia de
Materiales de Madrid, CSIC, Cantoblanco, E-28049 Madrid, Spain}

\begin{abstract} We report detailed magnetization measurements on the
perovskite oxide NdNiO$_3$. This system has a first order
metal-insulator (M-I) transition at about 200 K which is associated with
charge ordering. There is also a concurrent paramagnetic to
antiferromagnetic spin ordering transition in the system. We show that
the antiferromagnetic state of the nickel sublattice is spin canted. We
also show that the concurrency of the charge ordering and spin ordering
transitions is seen only while warming up the system from low
temperature. The transitions are not concurrent while cooling the system
through the M-I transition temperature. This is explained based on the
fact that the charge ordering transition is first order while the spin
ordering transition is continuous. In the magnetically ordered state the
system exhibits ZFC-FC irreversibility, as well as history-dependent
magnetization and aging. Our analysis rules out the possibility of
spin-glass or superparamagnetism and suggests that the irreversibility
arises from magnetocrystalline anisotropy and domain wall pinning.
\end{abstract}



\maketitle

\section{Introduction} The rare earth nickelates (RNiO$_3$, R $\neq $
La) have been under active investigation for the past two decades
because of the interesting electronic and magnetic properties exhibited
by these systems.\cite{Medarde, Catalan} These oxides undergo a
bandwidth controlled metal-insulator (MI) transition on changing the
temperature, chemical or hydrostatic pressure.\cite{Torrance, Obradors,
Canfield, Zhou, Ashutosh} In the metallic state the structure of these
nickelates is that of an orthorhombic distorted perovskite with space
group ${Pbnm}$.\cite{P.Lcorre} The metal to insulator transition occurs
with a structural transition which consists of an increase in the unit
cell volume, a decrease in Ni-O-Ni bond angle and a symmetry lowering
from orthorhombic ${Pbnm}$ to monoclinic $P2_{1}/n$. The symmetry
lowering is understood in terms of charge ordering with a charge
disproportionation 2Ni$^{3+}\to$ Ni$^{3+\delta}+$ Ni$^{3-\delta}$ with
$\delta \approx 0.2-0.3$.\cite{Alonso_1, Alonso_2, Alonso_3, Alonso_4,
Staub, Medarde1} In the early reports, the M-I transition of these
compounds was attributed to the opening of an Ni-O charge transfer gap
created by band narrowing.\cite{Torrance} But the occurrence of charge
ordering at the M-I transition and some recent theoretical calculations
suggest that the M-I transition owes its origin to the opening of a gap
between the spin up $e_\textnormal{g}$ band of Ni$^{3-\delta}$ and the hardly spin
polarized $e_\textnormal{g}$ band of Ni$^{3+\delta}$.\cite{Mazin} In these compounds
the higher temperature phase is metallic and the lower temperature phase
is insulating. The M-I transition is of first order and is associated
with a large thermal hysteresis and time dependent effects in transport
properties such as resistivity and thermopower.\cite{Granados, Devendra,
Devendra1, Devendra2} During the cooling process, in the temperature
window where hysteresis is seen, these compounds phase separate into
insulating and supercooled metallic regions. The supercooled regions are
metastable and they switch over to the insulating state stochastically
giving rise to time dependence and hysteresis in transport
properties.\cite{Devendra, Devendra1, Devendra2}

The nickelates also undergo a temperature driven magnetic transition,
which is relatively less studied, because the higher magnetic moment of
rare earth ion (e.g. Nd$^{3+}$ moment $\approx 3.6\mu_\textnormal{B}$) makes it
difficult to get any information about the magnetic ordering of the Ni
sublattice (Ni$^{3+}$ moment $\approx 1 \mu_\textnormal{B}$) through magnetization
measurements.\cite{Blasco, Perez, Medarde} Muon spin rotation
experiments of Torrance et al. show that these compounds undergo a
magnetic ordering from paramagnetic to an antiferromagnetic state on
lowering the temperature.\cite{Torrance} The magnetic ordering
temperature ($T_{\textnormal{N}}$) coincides with the M-I transition
temperature ($T_\textnormal{MI}$) for PrNiO$_3$ and NdNiO$_3$, while it
is lower than $T_\textnormal{MI}$ for all the other nickelates. The
magnetic transition is of second order for all nickelates having
$T_\textnormal{MI} > T_\textnormal{N}$, \cite{Perez1, Caytuero} but for
NdNiO$_3$ and PrNiO$_3$ where $T_\textnormal{MI}=T_\textnormal{N}$, the
nature of the magnetic transition is difficult to probe independently.
While one would expect the magnetic transition to be continuous as
seen in other members of the series we note that there is at least one report
which goes against this expectation and claim that the said transition is of
first order.\cite{Caytuero}

Neutron diffraction experiments show that, below $T_\textnormal{N}$, the
magnetic arrangement of Ni moments is characterized by the propagation
vector ($\frac{1}{2}$,0,$\frac{1}{2}$) which suggests three possible
magnetic structures, of which, two are collinear and one is
non-collinear.\cite{Garc, Garc1, Alonso_1, Casais, Giovannetti} The
collinear magnetic structure consists of up-up down-down stacking of Ni
magnetic moments, where each Ni moment is antiferromagnetically coupled
to three of its nearest neighbors and ferromagnetically to the remaining
three nearest neighbours. This magnetic structure implies that the
orbital degeneracy of Ni$^{3+}$ $e_\textnormal{g}^1$ electrons should be lifted by an
orbital ordering, a prediction which has not gathered any experimental
support so far.\cite{Garc1} Soft X-ray resonant scattering experiments
at the Ni $L_{ 2,3}$ edges show that the ($\frac{1}{2}$,0,$\frac{1}{2}$)
reflections are purely of magnetic origin with no orbital contribution
whatsoever thus more or less ruling out collinear magnetic order in the
system.\cite{Scagnoli} In fact, the orbital degeneracy of the Ni$^{3+}$
$e_\textnormal{g}^1$ electron is found to be lifted by
charge-ordering\cite{Mazin} and this supports the existence of a
non-collinear magnetic structure which does not require orbital
ordering. The low temperature specific heat
data and the resonant soft X-ray diffraction data of induced Nd magnetic
moment in NdNiO$_3$ indicate that, in all likelihood, the ordering of Ni
moments in NdNiO$_3$ is non-collinear.\cite{Bartolome, Scagnoli1}

In this work, for the first time, we report the magnetization of the Ni
\emph{sublattice}, which we extracted after carefully subtracting the
contribution of the Nd moments from the total magnetization. The
magnetization of the Ni sublattice shows weak ferromagnetism which
indicates that the magnetic arrangement of the Ni moments is perhaps
canted. The existence of weak ferromagnetism cannot be understood in
terms of the magnetic structures referred to in the previous paragraph,
even the noncollinear ones. This suggests that those magnetic structures
do not represent the true picture and the actual magnetic arrangement of
Ni moments could be quite different from what has been thought of so
far. Further, we found that the supercooled metallic phase is
magnetically ordered which indicates that the transition, on cooling,
from the paramagnetic to the antiferromagnetic state happens at the
nominal transition temperature ($\approx${~200~K}) unlike the
metal-insulator transition which is broadened and happens at lower
temperatures as the supercooled metallic regions switch to the
insulating phase stochastically. This shows that the
connection between the magnetic transition and the metal-insulator
transition is rather weak and they do decouple if the system is
supercooled.
%\textcolor{red}{The absence of thermal hysteresis in the
%magnetic transition suggests that the transition is continuous.}
Also, the magnetization of the Ni sublattice shows features such as FC-ZFC
irreversibility which is indicative of the presence of frustration in
the weak ferromagnetic state.

\section{Experimental Details} High quality polycrystalline NdNiO$_3$
pellets were prepared by a liquid mixture technique described
elsewhere.\cite{Massa}

All the magnetic measurements were performed in a SQUID magnetometer
(Quantum Design, MPMS XL). Since, in this work, we are trying to extract
the small signal from the Ni moments buried under the much larger signal
from the Nd moments it is a sine qua non that we are absolutely sure
about the quality of the data. The magnetic signal from the samples of
NdNiO$_3$ and NdGaO$_3$, each of mass about 120mg, is 0.00262~emu and
0.00159~emu respectively at 150~K and 500~G. These numbers are more than
three orders of magnitude higher than the level where artifacts start
distorting the measured data.\cite{Ney} Further, the sample holders used
in SQUID measurements can give rise to misleading results when the
background signal from the sample holder becomes large enough so
that it can no
longer be ignored compared to the signal from the sample.\cite{Casa} In
our case, the sample holder is a piece of straw which gives a
temperature independent signal of about $-4\times 10^{-6}$ emu at 500~G
which is about 600 times smaller than the signal from the NdNiO$_3$
sample at 150~K. From the aforementioned we see that artifacts or
extraneous contributions are negligible compared to the magnetic signal
of NdNiO$_3$, and thus, our SQUID data can be confidently used for the
critical analysis we are setting out to do.

The field dependent resistivity measurements were performed in a home
made cryostat placed between the pole pieces of a large electromagnet.
More details on the resistivity measurements are available in one of our
earlier publications.\cite{Devendra}

% figure 1
\begin{figure} \begin{centering}
\includegraphics[width=0.8\columnwidth]{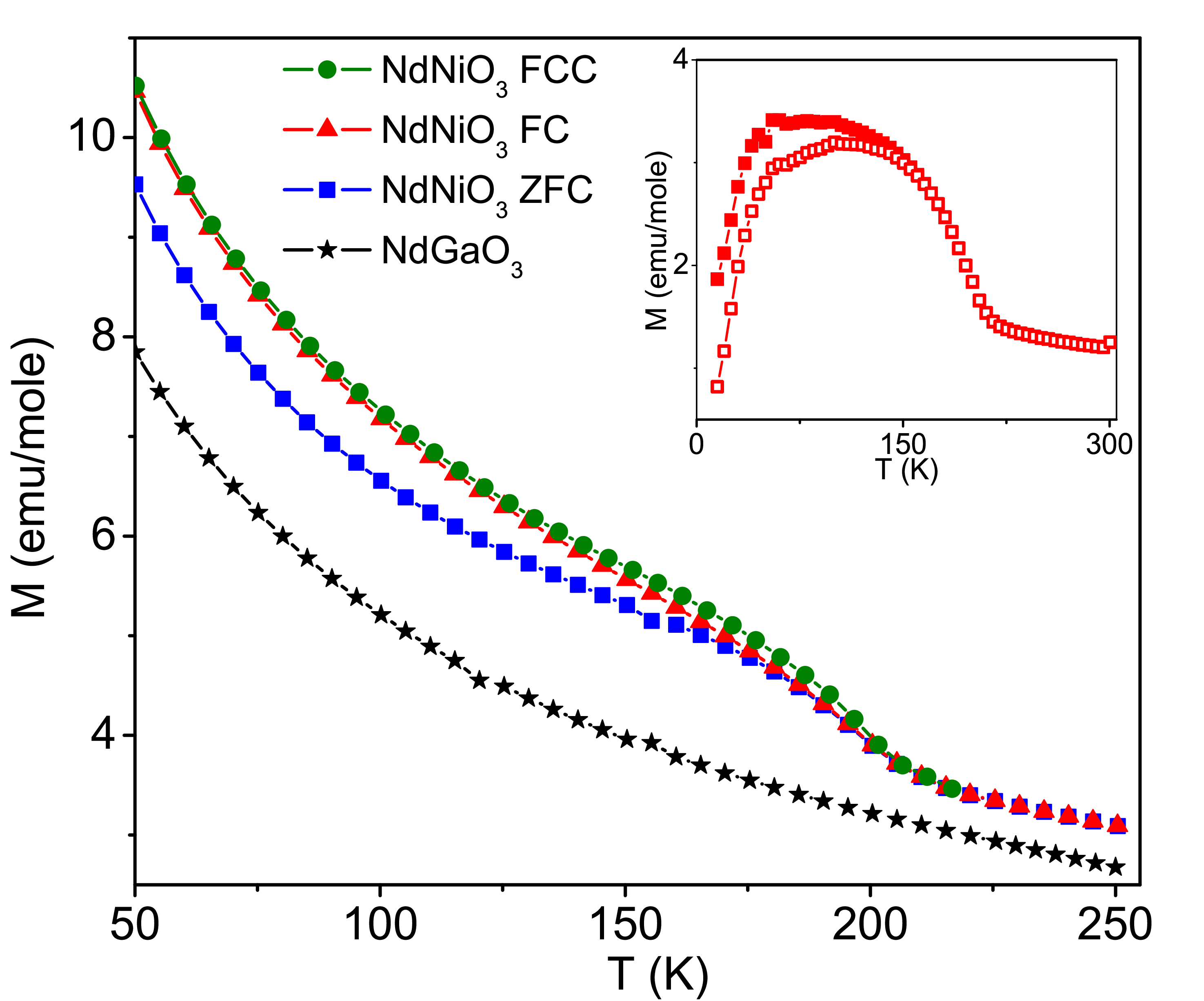} \par\end{centering}
\caption{(Color Online) The temperature variation of the magnetization
of NdNiO$_3$ in FCC (circles), FC (triangles), and in ZFC (squares)
protocols at 500 Oe. The stars show the magnetization of NdGaO$_3$
at the same field. For NdGaO$_3$ the FCC, FC and ZFC magnetizations
coincide. The inset shows the difference in magnetization of NdNiO$_3$
and NdGaO$_3$ down to 10~K at 1000 Oe in FC (upper curve, filled
squares) and ZFC (lower curve, open squares) protocol. We used 119~mg of
NdNiO$_3$ and 118~mg of NdGaO$_3$ for these measurements.} \label{fig: M
raw} \end{figure}

%figure 2
\begin{figure} [t] \begin{centering}
\includegraphics[width=1.0\columnwidth]{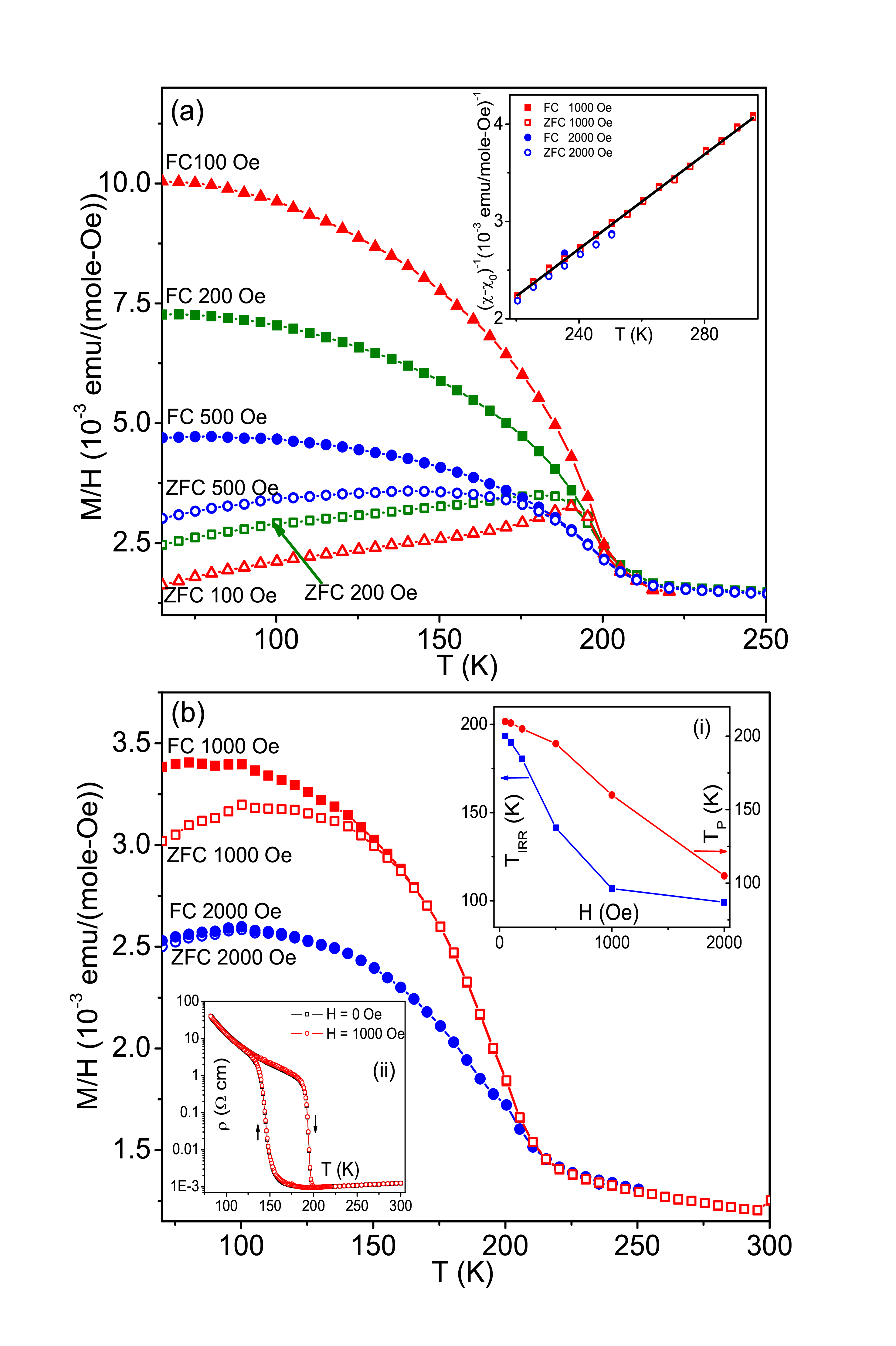}
\par\end{centering} \caption{(Color Online) The dc magnetic
susceptibility of the Ni sublattice versus temperature for FC and ZFC
protocols at various fields. The inset of (a) shows that the
susceptibility above 220~K follows the modified Curie-Weiss law shown in
equation (\ref{eq:Curie Weiss}) quite closely. The top-right inset of
(b) shows the temperature dependence of resistivity at zero field and
1000 Oe during cooling as well as warming. The bottom-left inset of (b)
shows how $T_\textnormal{IRR}$ and $T_\textnormal{P}$ depend on the
applied field. $T_\textnormal{P}$ is determined by Gaussian fitting of
the ZFC curves close to their maxima. }
\label{fig: M vs T} \end{figure}

%figure 3
\begin{figure}[h] \begin{centering}
\includegraphics[width=0.8\columnwidth]{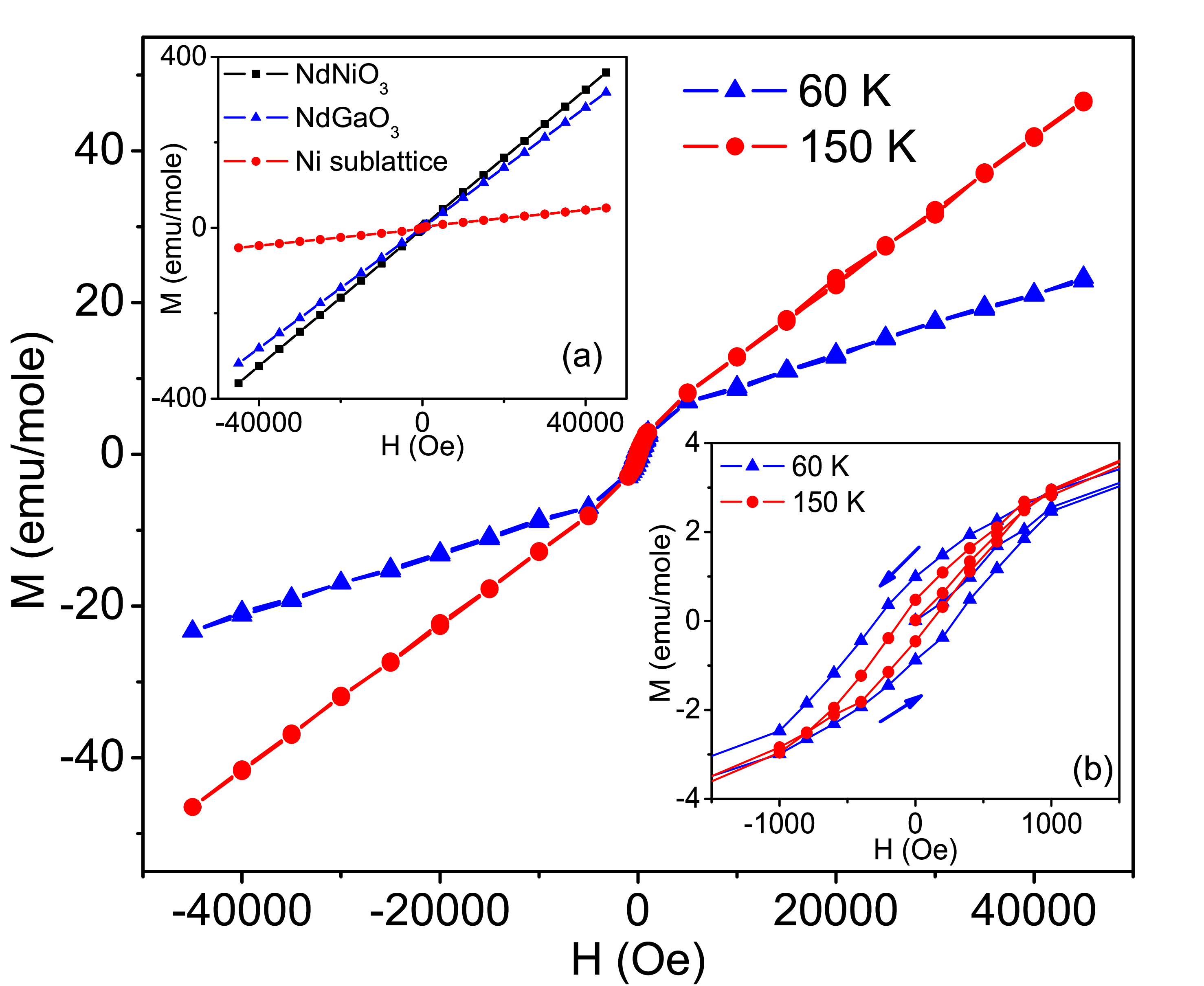} \par\end{centering}
\caption{(Color Online) M-H curves for the Ni sublattice at 150~K and
60~K. The inset (a) displays the magnetization of NdNiO$_3$, NdGaO$_3$,
and their difference at 150~K. The inset (b) shows an expanded view of
the low field data of Ni sublattice.} \label{fig: M vs H} \end{figure}

\section{Results and Discussion} \subsubsection{Magnetization
measurements} Figure \ref{fig: M raw} shows the magnetization of
NdNiO$_3$ and NdGaO$_3$ in FC, ZFC and FCC protocols at 500 Oe. In the
FC protocol we cool the sample in the presence of a specified field and
then record the magnetization while slowly warming up the sample keeping
the field fixed. In the ZFC protocol we cool the sample in zero field to
the lowest temperature and then apply the specified field and record the
magnetization while warming up. In FCC protocol the magnetization is
recorded while cooling in the specified field. The magnetization plots
of NdNiO$_3$ show a shoulder around 200~K attributable to the
ordering of Ni moments. We see that below 200~K  the magnetization
of NdNiO$_3$ depends on the experimental protocol. The FCC magnetization
is slightly higher than the FC magnetization while ZFC magnetization is
lower than both FCC and FC magnetizations. Above 200~K, the FCC, FC, and
ZFC curves overlap and are indistinguishable. The existence of thermal
and magnetic history dependence in magnetization suggests that the
system is not in thermodynamic equilibrium below 200~K. In contrast, for
the reference sample NdGaO$_3$, the magnetization values in FCC, FC, and
ZFC protocol coincide and follow the Curie law.

\begin{table} \begin{centering} \begin{tabular}{|c|c|c|c|c|c|} \hline\
Field (Oe) & C & $\theta$ & $\chi_0$ & $\chi^{2}/DOF$ & $R^{2}$
\tabularnewline \hline\ 1000(FC) & 0.043(4) & 125(6) & 0.00095(2) &
1.314 & 0.99916 \tabularnewline \hline\ 1000(ZFC) & 0.043(4) & 126(6) &
0.00095(2) & 1.271 & 0.99920 \tabularnewline \hline \end{tabular}
\par\end{centering} \caption{Fit parameters obtained from the fitting of
equation (\ref{eq:Curie Weiss}) to the 1000 Oe magnetic susceptibility data
of figure \ref{fig: M vs T} above 220~K. The quality of the fit is clear from
the fitted line to the red squares in the inset of fig \ref{fig: M vs T}(a) as
well as from the low $\chi^{2}/DOF$ values and the $R^2$ values very close to unity presented in this table. For other
field values the number of data points above 220~K and their span are
not good enough to warrant comparable quality of fitting.
\label{tab:fitting-parameters}} \end{table}

To extract the magnetization of Ni sublattice from the experimental data
we subtract the contribution of Nd moments from that of NdNiO$_3$. The
Nd magnetic moment is estimated from the magnetization data of NdGaO$_3$
which has the same crystal structure and almost the same lattice
parameters as NdNiO$_3$.\cite{Devendra2} Since gallium and oxygen ions
have no magnetic moment, the magnetization of NdGaO$_3$ arises only from
the contributions of the Nd moments sitting at the A sites of the
perovskite structure. By subtracting the NdGaO$_3$ magnetization (per
mole) from that of NdNiO$_3$ we should be able to calculate the
magnetization of Ni sublattice, provided Nd moments behave in the same
fashion in both NdGaO$_3$ and NdNiO$_3$. Unfortunately this method runs
into rough weather because the Nd moments in NdNiO$_3$ tend to order at
low temperature aided by the ordering of the Ni sublattice.

Neutron diffraction measurements on bulk NdNiO$_3$ show that the
magnetic ordering of Nd moments starts below 40~K,\cite{Garc1, Garc2}
while the synchrotron radiation data on thin films of NdNiO$_3$ suggest
that magnetic ordering of Nd moments starts at T$_\textnormal{MI}$ but
becomes significant only at low temperatures below 70 K.\cite{Scagnoli1}
The higher Nd ordering temperature seen in the thin films may have a
possible connection with the epitaxial strain in the films.\cite{Baena}
The ordering of Nd moments is thought to be induced by the direct
exchange interaction with the neighboring Ni moments and is
antiferromagnetic in nature while the Nd moments in NdGaO$_3$ remain
paramagnetic throughout the temperature range (See Ref.\
\onlinecite{Scagnoli1} and Fig. \ref{fig: M raw}). So on cooling below
the magnetic ordering temperature of Nd, the difference in the
magnetization of NdNiO$_3$ and NdGaO$_3$ would drop drastically because
the contribution of Nd moments to the magetization of NdNiO$_3$ would
fall due to their antiferromagnetic ordering. In our case, such a
drastic drop in the difference in magnetization of NdNiO$_3$ and
NdGaO$_3$ is seen to occur below about 50~K as is clear from the inset
of Fig. \ref{fig: M raw}. This suggests that the effect of Nd ordering
becomes quite significant below 50~K, and sufficiently above this
temperature, the magnetization of Ni sublattice could be obtained, to a
reasonable degree of confidence, by the subtraction of NdGaO$_3$
magnetization from that of NdNiO$_3$.

\subsubsection{Magnetic ordering of the Ni sublattice} In figure
\ref{fig: M vs T} we show the temperature dependence of ZFC and FC dc
magnetic susceptibility of Ni sublattice between 100~Oe to 2000~Oe.
Above 220~K, as is clear from the inset of figure  \ref{fig: M vs T}(a), the data fit well to the modified Curie Weiss equation
\begin{equation} \chi = C/ (T-\theta) + \chi_{0} \label{eq:Curie Weiss}
\end{equation} where $C$ and $\theta$ are Curie and Weiss constants
respectively, and $\chi_{0}$ is a constant arising from Van Vleck and
Pauli paramagnetism and Landau and core diamagnetism. The parameters obtained from the fitting of equation
(\ref{eq:Curie Weiss}) to the 1000~Oe susceptibility data of figure
\ref{fig: M vs T}, in the temperature range of 220-300~K, is shown in
table \ref{tab:fitting-parameters}. The $R^2$ values very close to unity and the low $\chi^{2}/DOF$ values indicate that fit quality is very good.

The presence of possible defects in the crystalline lattice structure
may also give a contribution to magnetic susceptibility, but that
contribution is generally around $10^3$ times smaller than our measured
signal,\cite{Nair, Bornemann} and this fact allows us to ignore them.
%The Pauli and Landau contributions arise from itinerant electrons.

The subtraction of NdGaO$_3$ magnetic susceptibility from that of
NdNiO$_3$ cancels the temperature independent Van-Vleck and core
contribution of Nd ions, and so $\chi_{0}$ is free of these two. The
core diamagnetic susceptibility of Ni ions is around $-68\times10^{-6}$
emu/mole\cite{Xu} and the Landau diamagnetic susceptibility is connected
to Pauli paramagnetic susceptibility by the equation $\chi^{Landau} =
-(1/3)[m/m^*]^2\chi^{Pauli}$, where $m$ is the free electron mass and
$m^*$ is the effective mass of an electron in the conduction band. Since
$m^*$ is found to be significantly larger than $m$ in this family of oxides,
\cite{Rajeev} the
$\chi^{Landau}$ can be neglected in comparison to
$\chi^{Pauli}$.\cite{Xu} Thus the $\chi_{0}$ values shown in table
\ref{tab:fitting-parameters} arise predominantly  from the Pauli
paramagentism of itinerant electrons, and they are in good agreement
with the values reported in Refs.\ \onlinecite{Perez} and \
\onlinecite{Xu}. The Pauli paramagnetic susceptibility of NdNiO$_3$ is
around two orders of magnitude larger than that calculated using the
free-electron value which suggests that the electron correlation in these
systems is very strong.\cite{Perez} It is to be noted that we get a positive Weiss
constant $\theta$ which is indicative of a ferromagnetic interaction in
the magnetically ordered state. This is surprising considering the fact
that neutron and resonant soft X-ray diffraction measurements show that
the system has antiferromagnetic order below
$T_\textnormal{MI}$.\cite{Garc, Garc1, Alonso_1, Casais, Giovannetti} In
consonance with the above observation of a positive Weiss constant
we point out  that below 195~K, in FC
measurements, the magnetic susceptibility increases on decreasing the
temperature as would be expected in the case of ferrimagnets or canted
antiferromagnets which behave as weak ferromagnets.
 See figure \ref{fig: M vs T}.
% figure 3

In figure \ref{fig: M vs H} we have shown the field dependence
of the magnetization of the Ni sublattice.
The inset (a) of the figure shows the magnetization versus
field for NdNiO$_3$, NdGaO$_3$ and the Ni sublattice at 150~K. The Ni
sublattice magnetization is obtained by subtracting
the contribution of Nd moments (obtained from NdGaO$_3$) from that of
NdNiO$_3$. In the main panel of figure \ref{fig: M vs H} and its inset (b) we
show the
magnetization versus field for the Ni sublattice at 150~K and 60~K. The
M-H curves show a small hysteresis at small fields, while at higher
fields, the M-H curves behave as that of a typical antiferromagnet, with
$M$ varying linearly with $H$, which leads to the conclusion that this
system is a spin-canted antiferromagnet.\cite{chikazumi} The presence of
spin canted magnetism (weak ferromagnetism) cannot be explained on the
basis of the magnetic structures proposed in the literature (Refs. 23,
24, 26). This is because the sum of the Ni magnetic moments in the
proposed collinear as well as the non-collinear magnetic structure is
zero (See figure 5 of Ref. 28). Thus our experimental data clearly show
that the magnetic structures proposed in the literature are not
the true magnetic picture of NdNiO$_3$.  Further investigations are required to
confirm this new experimental finding.

Referring to the inset (b) of figure \ref{fig: M vs H}, we see that, the
coercivity ($H_\textnormal{C}$) is temperature dependent below
$T_\textnormal{N}$, and it increases on lowering the temperature. Since
coercivity is related to magnetic anisotropy, this suggests that the
magnetic anisotropy increases on decreasing the temperature.

% figure 4
\begin{figure}[!t] \begin{centering}
\includegraphics[width=1.0\columnwidth]{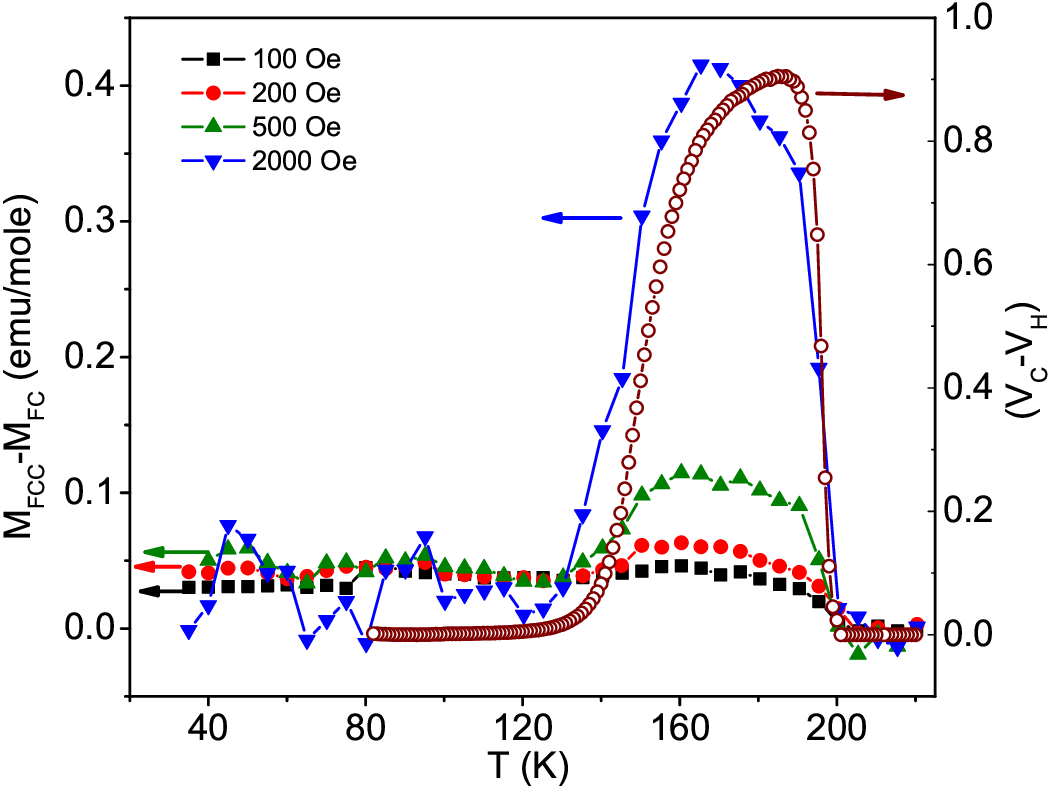} \par\end{centering}
\caption{(Color Online) The temperature variation of the difference in
magnetization, $M_\textnormal{FCC} - M_\textnormal{FC}$, of NdNiO$_3$
between cooling and heating runs at 100~Oe, 200~Oe, 500~Oe and 2000~Oe
(solid symbols). The open circles show the difference in the metallic
volume fraction, $V_\textnormal{C}-V_\textnormal{H}$, between cooling
and heating runs.} \label{fig: M diff} \end{figure}

\subsubsection{Magnetic state of the supercooled phase}
The transport properties of NdNiO$_3$ show thermal
hysteresis which is attributed to the presence of supercooled metallic
regions below the transition temperature.\cite{Devendra, Devendra1,
Devendra2} Now the question we would like to ask is this: What is the
magnetic state of the supercooled metallic regions?  Are they paramagnetic or
antiferromagnetic? In other words we are asking whether the paramagnetic
to antiferromagnetic transition,  when we cool the system through its magnetic
transition temperature (200~K ), takes place at that temperature or does  it
take place along with the M-I transition of the metastable phase at a lower
temperature? In order to throw some light on this issue we
measured the thermal hysteresis of magnetization. In figure \ref{fig: M
diff} we show the difference in cooling and heating cycle magnetization,
$M_\textnormal{FCC} - M_\textnormal{FC}$, of NdNiO$_3$ at a few field
values in the range 100 Oe to 2000 Oe. The data show that between 200~K
and 120~K, the magnetization of the cooling cycle is higher than that of
the heating cycle. The difference in the magnetization is maximum around
170~K. Figure \ref{fig: M diff} also shows the difference in the
metallic volume fractions between the cooling and heating runs
$V_\textnormal{C}-V_\textnormal{H}$, taken from reference
\onlinecite{Devendra}. The difference in the magnetizations and the
difference in the metallic volume fractions have remarkably similar
temperature dependence which suggests that they originate from a common underlying
physical mechanism. In a cooling run, below $T_\textnormal{MI}$, the
system contains supercooled metallic and insulating regions, while in a
heating run, it is mostly insulating.\cite{Devendra, Devendra1,
Devendra2} Therefore $V_\textnormal{C}-V_\textnormal{H}$ represents the
volume fraction of supercooled metallic regions. So the correlation
between the thermal hysteresis in magnetization and the supercooled
metallic volume fraction indicates that the supercooled metallic regions
have a higher magnetic moment compared to the insulating regions.

The Ni moments are paramagnetic in the normal metallic state
($T>T_\textnormal{MI}$) while they show a spin-canted antiferromagnetic
ordering in the insulating state. Also, the spin-canted insulating state
has a higher susceptibility than the paramagnetic metallic state (see
figure \ref{fig: M vs T}). This suggests that if the supercooled
metallic regions were paramagnetic, as above $T_\textnormal{MI}$, then
the magnetization of NdNiO$_3$ in a cooling run, where below
$T_\textnormal{MI}$ the system consists of supercooled metallic and
insulating regions, should be lower than that in a heating run where the
system is expected to be almost fully insulating. But the experimental
results discussed in the previous paragraph contradict this which
indicates that the supercooled metallic regions are not paramagnetic. To
make things more concrete, we compare the observed difference in the
magnetization of cooling and heating runs to the expected value of the
difference if the supercooled regions were paramagnetic. In the cooling
run, at 170~K, the volume fraction of the supercooled metallic regions
is around 0.9 from figure \ref{fig: M diff}. The dc magnetic
susceptibility of the paramagnetic metallic phase at 2000~Oe (Figure
\ref{fig: M vs T}), extrapolated down to 170~K, is about 20\% smaller
than that of the insulating phase which suggests that if the thermal
hysteresis in the magnetization is because of paramagnetic ordering of
supercooled metallic regions, then, according to our estimate, the
difference in the magnetization of the cooling and heating runs should
be around $-0.9$ emu/mole. But the observed difference in the
magnetization is $+0.4$ emu/mole which has the wrong sign and is smaller
in magnitude than the expected value. This observation strongly suggests
that the supercooled metallic regions are antiferromagnetic with canted
spins just like the insulating state. The small positive difference in
magnetization between cooling and heating is proportional to the volume
fraction of supercooled metallic regions and hence we conclude that this
difference in susceptibility is temperature independent. This suggests
that the observed difference in cooling and heating cycle magnetization
is coming from itinerant electrons in the supercooled metallic state through Pauli
paramagnetic and Landau diamagnetic contributions.\cite{endnote1} Thus
we see that the metallic state is paramagnetic above $T_\textnormal{MI}$
and on cooling below $T_\textnormal{MI}$, while a fraction of the high
temperature metallic phase exists in its supercooled state, the magnetic
ordering of the whole sample switches to an antiferromagnetic state at
$T_\textnormal{N}$.

From the above discussion, we conclude that in NdNiO$_3$, even
though the charge ordering and magnetic ordering occur at the same
 temperature (in equilibrium) they are not strongly
coupled and occur independently of each other.
Incidentally, we note that except in PrNiO$_3$ and NdNiO$_3$ of the
nickelate series, the two transitions occur at different temperatures
which supports the conclusion that the two transitions are only weakly
coupled. The antiferromagnetic order of the supercooled metallic
regions rules out the presence of any metastable magnetic phase associated
with the magnetic transition and suggests that the
magnetic transition is continuous in nature. This result
removes the ambiguity associated with the nature of the magnetic transition in
nickelates where $T_\textnormal{MI}$ = $T_\textnormal{N}$; the magnetic
transition is continuous which is consistent with the other members
of the series where $T_\textnormal{MI}$ > $T_\textnormal{N}$.

%figure 5
\begin{figure}[!t] \begin{centering}
\includegraphics[width=0.9\columnwidth]{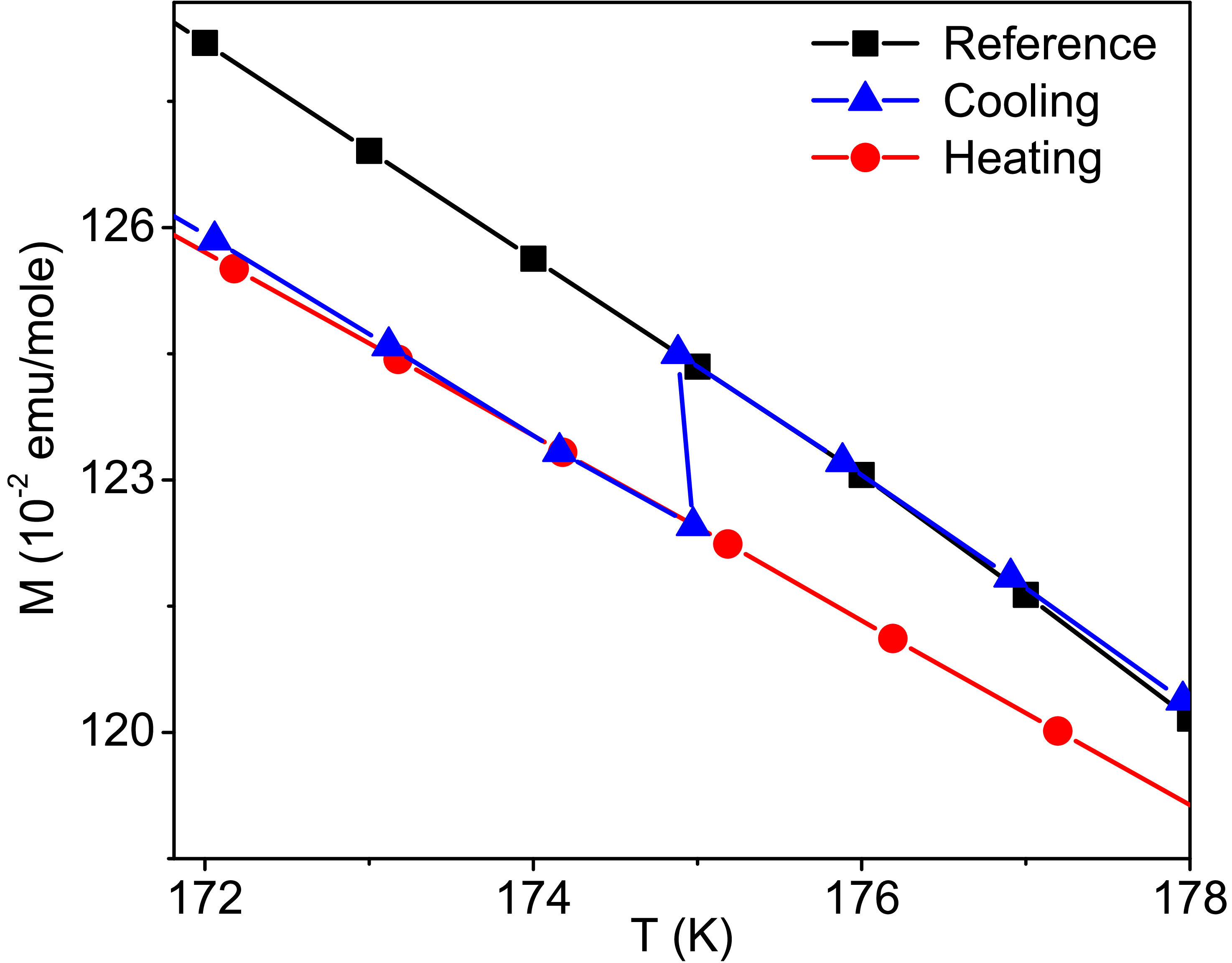} \par\end{centering}
\caption{(Color Online) Memory experiment in the FC protocol with
intermediate stops of one hour at 175, 150, 125 and 110~K. The field is
switched off during each stop. The data close to 175~K is shown here.
The black squares show the FC reference which is the magnetization in
FCC protocol(after removing the contribution of thermal hysteresis).}
\label{fig: FC memory} \end{figure}

%figure 6
\begin{figure}[] \begin{centering}
\includegraphics[width=0.8\columnwidth]{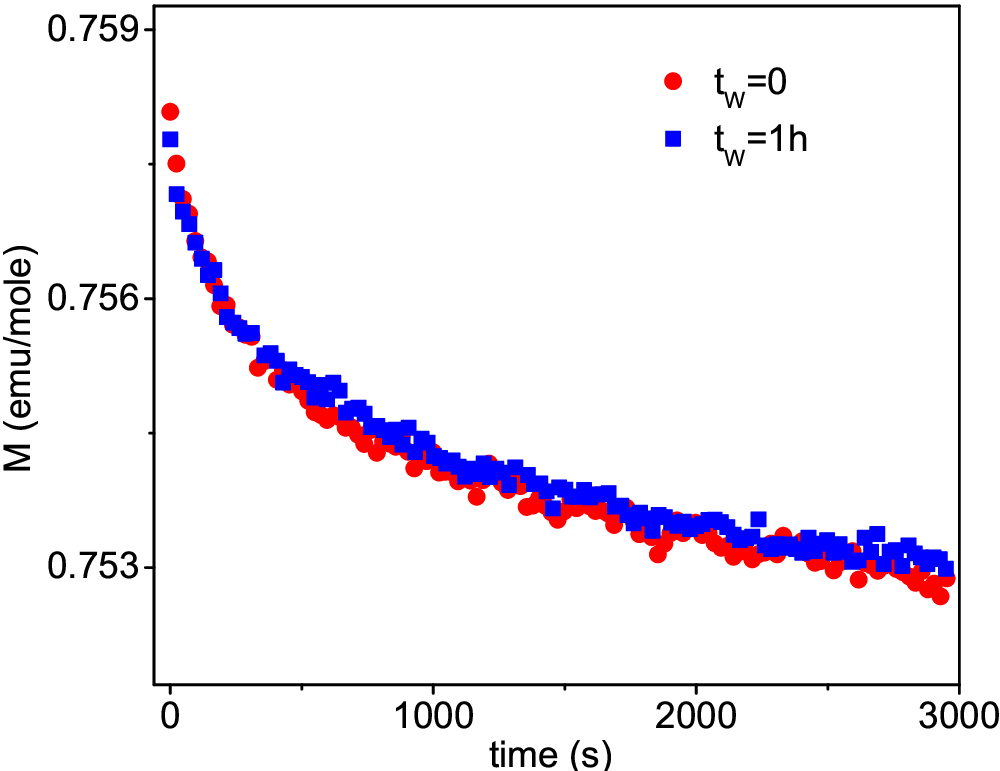} \par\end{centering}
\caption{(Color Online) The time decay of thermoremanent magnetization
of NdNiO$_3$ at 80~K (red circles). The blue squares show the decay of
thermoremanent magnetization with a one hour wait time.} \label{fig:
TRM} \end{figure}

\subsubsection{The FC-ZFC irreversibility} The FC and ZFC magnetic
susceptibilities show a history dependence with a bifurcation between
the two curves at a temperature known as the temperature of
irreversibility ($T_\textnormal{IRR}$). See inset (i) figure \ref{fig: M vs T}(b).
The temperature of irreversibility depends on the magnetic field and it
decreases on increasing the magnetic field. For fields greater than
2~kOe the FC and ZFC curves superpose. Behavior such as this where the
magnetic susceptibility depends on measurement history has been observed
in non-equilibrium systems such as spin-glasses,\cite{Maydosh, Tiwari,
Vijay} superparamagnets,\cite{Knobel} cluster-glasses,\cite{Deac, Huang}
supercooled systems,\cite{Chaddah, Chaddah1} and also in anisotropic ferromagnets
and ferrimagnets.\cite{Anil, Anil1, Anil2, Song, Roshkoa} The ZFC data
show a peak, and the peak broadens and shifts to low temperatures on
increasing the magnetic field. We analysed the nature of this peak and
found that the peak temperature ($T_\textnormal{P}$) as a function of
field ($H$) does not behave as in the case of spin-glasses,
cluster-glasses,\cite{Almeida} or superparamagnets\cite{Zheng, Vijay1}
which indicates that the system is neither a spin-glass nor a
superparamagnet. We also rule out supercooling as a possible reason for
the FC-ZFC irreversibility by the following argument. The resistivity
measurements show a thermal history dependence which is attributed to
the presence of supercooled metallic regions below $T_\textnormal{MI}$.
We did not observe any significant magnetic field or magnetic history
dependence in transport properties which suggests that the volume
fraction of supercooled metallic regions is not altered by the
application of a magnetic field. See inset (ii) of figure \ref{fig: M vs
T}(b). The lack of dependence of resistivity on applied magnetic field
has also been reported earlier by Mallik et al.\cite{Mallik} From these
results, we infer that the magnetic history dependence of the dc
magnetic susceptibility (see figure \ref{fig: M vs T}(a) and (b)) cannot
be originating from the supercooled metallic phases. So far our analysis
has shown that the magnetic hysteresis does not arise from spin-glass or
cluster-glass nature, superparamagnetism or supercooling. This leaves us with
the only possibility that the magnetic hysteresis in this system is
arising from magnetic anisotropy of the spin canted magnetic domains.

% figure 7
\begin{figure*}[!t] \begin{centering}
\includegraphics[width=2.0\columnwidth]{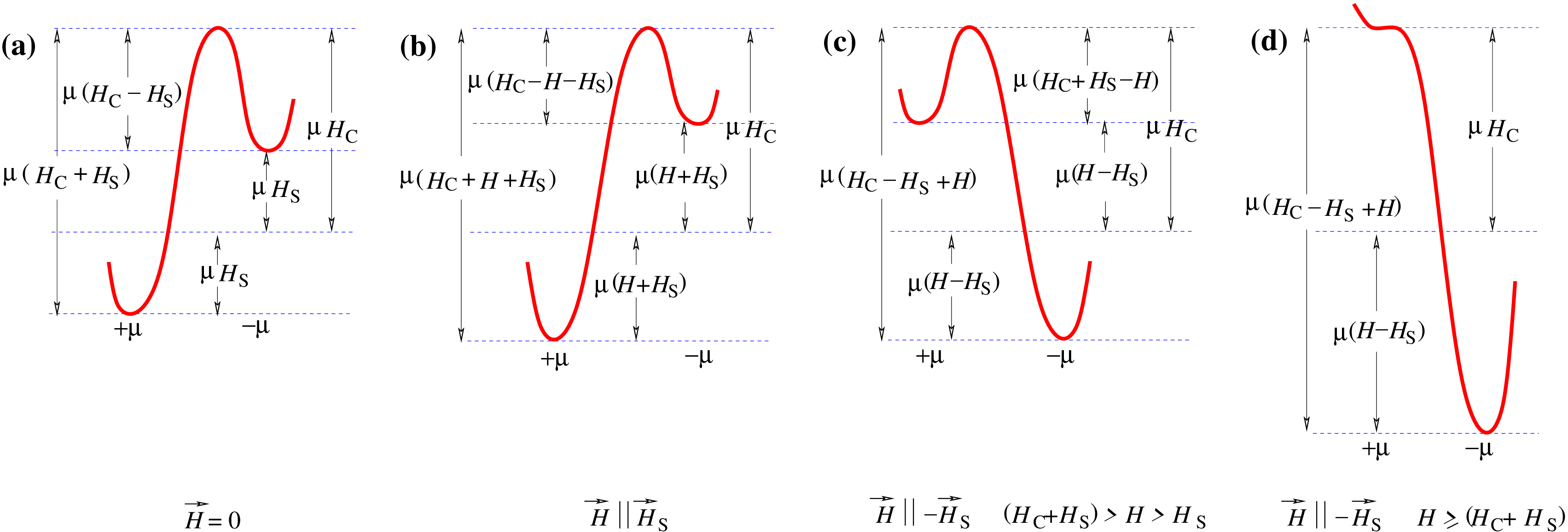} \par\end{centering}
\caption{(Color Online) The free energy profile of a bistable subsystem
at various applied fields} \label{fig: energy} \end{figure*}

To be doubly sure that the history dependent FC and ZFC susceptibility
of the Ni sublattice has nothing to do with superparamagnetism or
spinglass nature, we performed FC, ZFC memory and aging experiments.
Since the Nd moments are paramagnetic, they would not have any role in
the memory and aging of NdNiO$_3$. Thus if any such effect is seen in
this system it would have to be attributed to the Ni sublattice. The FC
memory experiments were performed with intermediate stops of one hour at
175, 150, 125, and 100~K. In these experiments the system is cooled in a
100~Oe field from 220~K to 80~K and then heated back to 180~K to remove
the influence of supercooled metastable regions on dynamic behavior.
Subsequently the system is cooled from 180~K to 80~K with intermediate
stops of 1 hour at 175, 150, 125, and 100~K. The field was switched off
during the intermediate stops. The magnetization is recorded while
cooling and then during the subsequent heating. The FC memory data at
175~K is shown in figure \ref{fig: FC memory}. We can see that
immediately after an intermediate stop the magnetization does not go
back to its pre-stop value after switching on the field. In the
subsequent heating run, we did not find any memory of the intermediate
stops and this rules out the possibility of superparamagnetism or
spin-glass behavior in the system.\cite{Vijay} We also carried out ZFC
memory experiments on the system at 170~K and the result was negative.
This confirms the conclusions we arrived at from the FC memory
experiments and once again rules out a spin-glass state.\cite{Vijay}

In figure \ref{fig: TRM} we show the results of the FC ageing
experiment. In this experiment one essentially measures the time decay
of thermoremanent magnetization along with wait time dependence. To
begin with we cool the system from 250~K to 80~K in the presence of
100~Oe field, wait for the duration $t_\textnormal{w}$ at 80~K with the
field on, and then switch off the field and record the magnetization as
a function of time. It is clear from the figure that the system does not
show any noticeable wait time dependence in FC ageing and this yet again
rules out the possibility of the system being a spin glass or a
superparamagnet.\cite{Vijay}

The irreversibility of the FC and ZFC magnetic susceptibility in a
system which is neither a spin-glass nor superparamagnetic can be
understood in terms of a competition between the magnetocrystalline
anisotropy and domain wall pinning on the one hand and applied field and
thermal energy on the other.\cite{Anil, Anil1, Anil2, Song, Roshkoa}
Below the temperature of magnetic ordering, a magnetically ordered
material consists of uniformly magnetized regions which are known as
magnetic domains. At any temperature $T$ and applied field $H$, the free
energy of the magnetic systems have a number of local minima which are
determined by the arrangement of the domains inside the magnetic
material. These local minima states are separated by energy barriers
which arise due to magnetocrystalline anisotropy and domain wall
pinning. When the thermal energy is greater than the energy barrier of
the metastable state in which the system is trapped, the system can
explore the neighboring states in search of the global minimum or the
equilibrium state. The free energy configuration is a function of
applied magnetic field $H$ and temperature $T$ and on changing $H$ or
$T$ (which changes the magnetocrystalline anisotropy) the system evolves
from one configuration to another.\cite{Bertotti} We shall make an
attempt to understand our system on the basis of the Preisach model in
which the free energy configuration is decomposed into an ensemble of
bistable subsystems.\cite{Bertotti} A bistable subsystem consists of two
metastable states separated by an energy barrier. The two states have
moments oriented in opposite directions and are termed as $\pm \mu$
states. The free energy of these states in the absence of applied
magnetic field is determined by the local interaction field
($H_\textnormal{S}$) and the the coercive field ($H_\textnormal{C}$).
$H_\textnormal{S}$ is the net magnetic field produced at the location of
the moment $\mu$ by the magnetic moments of all the neighboring domains.
If $H_\textnormal{S}=0$ then $\mu H_\textnormal{C}$ represents the
anisotropy energy barrier that has to be crossed to go from $+\mu$ to
$-\mu$ state or vice versa. The barrier height seen from the $+\mu$ side
is $\mu(H_\textnormal{C}+ H_\textnormal{S})$ while from the $-\mu$ side
it is $\mu(H_\textnormal{C}- H_\textnormal{S})$. See figure \ref{fig:
energy} (a). The application of a magnetic field ($H$) changes the free
energy of the metastable states which in turn affects the effective
height of the energy barrier. We also note that a change in the
temperature can also affect the free energy barrier through its effect
on magnetocrystalline anisotropy. \cite{Song, Roshkoa, Bertotti}

In the following paragraphs we discuss qualitatively the FC-ZFC
irreversibility and the remanent magnetization using the standard
Preisach model. Thereafter we apply it to understand the observed
results of aging experiments.

In ZFC protocol when the system is cooled below $T_\textnormal{N}$ each
subsystem will be in its lower energy state which is determined by
$H_\textnormal{S}$ (Figure~\ref{fig: energy}~(a)). On applying a
magnetic field, depending on the direction and strength of the applied
field, the low energy state of the subsystem may remain as the low
energy state (Figure~\ref{fig: energy}~(b)), or may become metastable or
unstable (Figure~\ref{fig: energy}~(c)~and~(d)). If $\vec{H} \parallel
-\vec {H_\textnormal{S}}$, the subsystems for which $H$ is larger than
$H_\textnormal{C}+H_\textnormal{S}$, will flip to their new low energy
state (Figure~\ref{fig: energy}~(d)). It is this flipping that gives
rise to the initial value of the ZFC magnetization of the system. The
subsystems for which $H$ is less than $(H_\textnormal{C} +
H_\textnormal{S})$, are now in a metastable state (Figure~\ref{fig:
energy}~(c)). These subsystems will undergo a thermally activated
transformation, which gives rise to a slowly rising time dependent ZFC
magnetization even if the magnetic field is held fixed. On increasing
the temperature, $H_\textnormal{C}$ decreases and because of this more
number of subsystems will flip to their new low energy state and this
increases the ZFC magnetization further. As one increases the
temperature the ZFC magnetization curve will attain a peak when the most
probable $H_\textnormal{C}$ value of the Barkhausen moment ($\mu$)
becomes equal to the applied field $H$.

In the FC protocol the subsystems get trapped in their low energy
states, as the sample is cooled through the magnetic ordering
temperature in the presence of an applied field. At a constant field, a
decrease in temperature increases the energy barrier (because of
increase in $H_\textnormal{C}$), but this does not affect the relative
positions of the $+\mu$ and $-\mu$ states. Thus in the FC protocol there
is hardly any change of state of the bistable subsystems when cooling
through $T_N$. The temperature dependence observed in the FC
magnetization is because of temperature dependence of the Barkhausen
moment $\mu(T)$. That is why the shape of an FC magnetization curve is
nearly the same for all fields.

If we switch off the applied field in the FC protocol, the subsystems
for which applied field $\vec{H}$ is opposite and greater in magnitude
than $\vec{H_\textnormal{S}}$ will result in their low energy state
becoming a high energy state and vice versa. This can be understood
looking at figure \ref{fig: energy} where the initial states shown in
figures \ref{fig: energy} (c) or (d) switch to the final state shown in
figure \ref{fig: energy} (a) on removal of the applied field. Of these
subsystems, those which have $H_\textnormal{S}\geq H_\textnormal{C}$,
will become unstable on removing the field, and their change of state
constitutes the initial loss of FC magnetization. The other subsystems
(which have $H_\textnormal{S}<H_\textnormal{C}$) will become metastable
and their thermally activated transformation from a metastable to a new
lower energy state gives a further slow decay in FC magnetization.

At this point let us examine the effect of aging (wait time dependence)
on the system. All the subsystems occupy their lower energy state on
cooling through $T_\textnormal{N}$. Thus, after cooling, if we wait for
a few hours before switching off (or on) the field, it will not affect
the population of the $\pm \mu$ states and hence we would not get any
effect of aging on magnetic relaxation.

\section{Conclusion} We performed detailed magnetization measurements on
NdNiO$_3$ and extracted the magnetization of Ni sublattice after
removing the contribution of the rare earth Nd ion. Our results indicate
the presence of weak ferromagnetism coexisting with antiferromagnetic
order in the Ni sublattice. We argued that the weak ferromagnetism is
due to canting of antiferromagnetic spins. Further we found that in
contrast to the normal metallic state, the supercooled metallic regions
are magnetically ordered. This shows that while cooling the metal insulator
transition occurs over a temperature range of 200~K to 110~K, the
magnetic ordering is sharp and occurs at 200~K. The absence of metastable
phases in the magnetic transition suggests
that the magnetic transition is continuous similar to other members
of the series that have $T_\textnormal{MI}$ > $T_\textnormal{N}$. Below
$T_\textnormal{N}$, the ZFC-FC magnetizations diverge exhibiting
irreversibilities, that could remind one of a spin-glass state. However,
our analysis shows that the system is neither a spin-glass nor a
superparamagnet, and the irreversibilities arise from the
temperature-dependent magnetocrystalline anisotropy and domain-wall
pinning.

\section{Acknowledgements} DK thanks the University Grants Commission of
India for financial support. JAA and MJM-L acknowledge the Spanish
Ministry of Education for funding the Project MAT2010-16404.
 \end{document}